\documentclass[%
reprint,showpacs,showkeys,amsmath,amssymb,
 aps,
]{revtex4-1}
\usepackage{amsmath}
\usepackage{cases}
\usepackage{amssymb}
\usepackage{CJK}
\usepackage{graphicx}
\usepackage{dcolumn}
\usepackage{bm}
\usepackage{color}
\usepackage[pagewise]{lineno}
\usepackage{subfigure}

\begin{document}
\begin{CJK*}{GBK}{} 

\preprint{APS/123-QED}

\title{A Bayesian-Neural-Network Prediction for Fragment Production in Proton Induced Spallation Reaction}

\author{Chun-Wang Ma
$^{1,2}$}
\thanks{Corresponding author. Email: machunwang@126.com}
\author{Dan Peng
$^{2}$}
\author{Hui-Ling Wei
$^{2}$}
\author{Yu-Ting Wang
$^{1,2}$}
\author{Jie Pu
$^{1,2}$}

\affiliation{$^{1}$
Institute of Particle and Nuclear Physics, Henan Normal University, \textit{Xinxiang 453007}, China \\
$^{2}$
School of Physics, Henan Normal University, \textit{Xinxiang 453007}, China \\
}

\begin{abstract}
Fragments productions in spallation reactions are key infrastructure data for various applications. Based on the empirical parameterizations {\sc spacs}, a Bayesian-neural-network (BNN) approach is established to predict the fragment cross sections in the proton induced spallation reactions. A systematic investigation have been performed for the measured proton induced spallation reactions of systems ranging from the intermediate to the heavy nuclei and the incident energy ranging from 168 MeV/u to 1500 MeV/u. By learning the residuals between the experimental measurements and the {\sc spacs} predictions, the BNN predicted results are in good agreement with the measured results. The established method is suggested to benefit the related researches in the nuclear astrophysics, nuclear radioactive beam source, accelerator driven systems, and proton therapy, etc.
\end{abstract}


\maketitle
\end{CJK*}

\section{Introduction}
Spallation reaction is one of the violent nuclear reactions, which happens when a high energy light particle hits on a target nucleus. The spallation reaction can naturally happen in cosmos where the high energy cosmic ray collides on nuclei \cite{Cosmic}, which results in the elements variation in universe. It can also artificially happen in nuclear facility, such as the accelerator driven systems (ADS) for nuclear waste disposal, radioactive nucleus production and so on, or during the proton therapy process using the accelerated protons. The incident energy of spallation reaction is above tens of $A$ MeV, which covers the range of the intermediate energy, the relativistic energy and even higher. As the result of spallation reaction, various of radioactive nuclei can be produced, the research of which has important applications in many disciplines including nuclear physics, nuclear astrophysics, the isotopic-separation-online (ISOL) type radioactive-ion-beam facilities, and the incoming third generation of radioactive nuclear beams facilities, the ADS for nuclear energy \cite{ADS2000} and nuclear waste transmutation \cite{ADS1999,INCL}, radioactive nuclei synthesis (especially for the extreme nuclei and nuclear isomers \cite{NIMApHf2009}), accelerator material radiation \cite{C14Acce}, proton therapy \cite{HT1,HT2}.

For its extensive applications, both experimental and theoretical interests have been attracted. In early time, the experiments were usually carried out using the accelerated proton on the synchrocyclotron or by the cosmic rays. After the reverse kinetic technique has been proposed, massive experiments have been performed to measure the spallation fragments above the $^{56}$Fe covering the incident energy from a few hundreds of $A$ MeV to above $A$ GeV. Prodigious data for fragments in spallation reactions have been assembled.

On the theoretical side, many models have been developed. The quantum molecular dynamics model has been improved for spallation reactions, for example,  \cite{IQMD-SJ19,IQMD-SJ20,IQMD-ZF19}. The statistical multi-fragmentation model \cite{SMM1995,SMM2001,SMM2005}, the Li\`{e}ge intranuclear cascade ({\sc incl++}) model \cite{INCL2013,INCL2014,INCL2015} (which has been implanted in the openMC, GEANT4 and FLUKA toolkits \cite{INCLFLUKA1,INCLGEANT4}), can be used to simulate the spallation reactions, which are usually followed by a secondary decaying simulation to reproduce the experimental results (A review is also recommended \cite{PPNPSpl19}).  Some semi-empirical parameterizations, including the {\sc epax} \cite{EPAX} and the {\sc spacs} \cite{spacs}, can globally predict the residue fragments in the spallation reaction. The nuclear energy agency (NEA) systematically compared the international codes and models for the intermediate energy activation yields to meet the needs in the ADS designation, energy amplification and medical therapy \cite{NEA}. Difficulties still exist for the reasons that the spallation reaction involves a wide range of incident energy, as well as a wide range of nucleus from the light to heavy one. Since the important applications of spallation reactions and the resultant residues productions, it is important to improve the theoretical predictions for the proton induced spallation reactions. The present models are limited in predicting the light fragments, in particular for the proton induced reaction at low energy. It is necessary to propose a new method for the exactly prediction of the fragments in spallation reactions.

The neural network method, as one kind of machine learning technologies, has found a rise of applications in nuclear physics. In the standard neural network, it is hard or even impossible to control the complexity of the model, which is likely to lead to overfitting problem and reduce the generalization ability of network \cite{FanCL}. The Bayesian neural network (BNN) provides a good manner to avoid overfitting automatically by defining vague priors for the hyperparameters that determines the model complexity \cite{MLP}. Prior distribution about the model parameters can be incorporated in Bayesian inference and combined with training data to control complexity of different parts of the model \cite{Jouko}. Successful examples of BNN applications in nuclear physics can be found in the predictions of nuclear mass \cite{Nmass1, Nmass2,Nmass3}, nuclear charge radii \cite{Nradii}, nuclear $\beta-$decay half-life \cite{Nhalf-life}, fissile fragments \cite{fission}, and spallation reactions \cite{spal}. Based on the vast numbers of measured fragments around the world-wide laboratories, it is quite promising to predict spallation cross sections accurately and give reasonable uncertainty evaluations with the BNN approach.

In this article, a new method is proposed to predict the fragment cross sections in proton induced spallation reactions. The BNN method is described in Sec. \ref{BNN}. The results and discussions are presented in Sec. \ref{RSTD}, and a conclusion is given in Sec. \ref{CONCLD}.

\section{BNN method}
\label{BNN}
The key principle of Bayesian learning is to deduce the posterior probability distributions through the prior distribution. The process of Bayesian learning is started by introducing the prior knowledge for model parameters. Based on the given training data $D = (x^{(i)},y^{(i)})$ and model assumptions, the prior distributions for all the model parameters are updated to the posterior distribution using the Bayes' rule,
\begin{equation}\label{E1}
p(\theta|D)=p(D|\theta)p(\theta)/p(D)\propto L(\theta)p(\theta).
\end{equation}
where $\theta$ denote the model parameters. The posterior distribution $p(\theta|D)$ combines the likelihood function $p(D|\theta)$ with the prior distribution $p(\theta)$, which contains the information about $\theta$ derived from the observation and the background knowledge, respectively.

The introduction of prior distribution is a crucial step which allows the prediction to go from a likelihood function to a probability distribution. The normalized quantity $p(D)$ can be directly understood as the edge distribution of the data, which can be obtained from the integration of the selected model hypothesis and prior distribution $p(\theta)$,
\begin{equation}\label{Eint}
p(D)=\int_{\theta}p(D|\theta)p(\theta)d{\theta}.
\end{equation}

In this work, the prior distributions $p(\theta)$ are set as the Gaussian distributions. The precisions (inverse of variances) of these Gaussian distributions are set as gamma distributions \cite{Nmass2}, which automatically control the complexity of different parts of the model. The likelihood function $p(D|\theta)$ and the objective function $\chi^{2}(\theta)$ are given by
\begin{eqnarray}
p(D|\theta)=exp(-\chi^{2}/2),    \nonumber\\
\chi^{2}=\sum\limits_{i}^{N} [y_{i}-f(x_{i};\theta)]^2 / \Delta{y_{i}^2}.   \nonumber
\end{eqnarray}
where $\Delta{y_{i}^2}$ is the associated noise scale. The function $f(x, \theta)$ is a multilayer perceptron (MLP) network, which is also known as the ``back-propagation'' or ``feed-forward''. A typical MLP network consists of a set of input variables ($x_{i}$), a certain hidden layers, and one or more outputs ($f_{k}(x; \theta)$). For an MLP network with one hidden layer and one output, the function is defined as,
\begin{equation}
f(x;\theta)= a+\sum_{j=1}^{H}b_j tanh(c_j+\sum_{i=1}^{I} d_{ji} x_i),
\end{equation}
where $H$ denotes the number of hidden unites, and $I$ is the number of input variables. $\theta = (d_{ji}, c_{j})$ and $\theta =(b_{j}, a)$ are the weights and bias of the hidden layers and output layer, respectively.

Based on the theoretical principles and prior knowledge, the posterior distribution be obtained from the data using Eq. (\ref{E1}). A predictive distribution of output $y^{new}$ for a new input $x^{new}$ is obtained by integrating the predictions of the model with respect to the posterior distribution of the model parameters,

\begin{equation}\label{posdis}
p(y^{new}|x^{new},D)=\int p(y^{new}|x^{new},\theta) p(\theta)d\theta.
\end{equation}

In the MLP network, what is interested is to get a reasonable prediction with the new input $x^{new}$ rather than the posterior distribution for parameters. For a new input $x^{new}$, the model prediction $y^{new}$ can be obtained from the mathematical expectation of posterior distribution,
\begin{equation}\label{E2}
y^{new}=E[y^{new}|x^{new},D]=\int f(x^{new},\theta) p(\theta|D)d\theta.
\end{equation}
The integral of Eq. (\ref{E2}) is complex, and a numerical approximation algorithm will reduce the complexity. The Markov chain Monte Carlo (MCMC) methods are applied to optimize the model control parameters, and obtain the predictive distribution. As one of the MCMC methods, the hybrid Monte Carlo (HMC) algorithm is firstly introduced by Neal \cite{Neal} to deal with the model parameters and Gibbs sampling for hyperparameters. The HMC is a form of the metropolis algorithm, where the candidate states are found by means of dynamical simulation. It makes the effective use of gradient information to reduce random walk behavior. In concept, the Gibbs sampler is the simplest Markov chain sampling methods, which is also known as the heatbath algorithm. The hyperparameters are updated separately using the Gibbs sampling, which allows their values to be used in chasing good step-sizes for the discretized dynamics, and helps to minimize the amount of tuning needed for a good performance in HMC. The integral of Eq. (\ref{E2}) is approximately calculated as,
\begin{equation}\label{E3}
y^{new}=1/K\sum\limits^{K}_{k=1}f(x^{new},\theta^{t}).
\end{equation}
where $K$ is the number of iteration samples. In a previous work, the BNN approach has been adopted to learn and predict the cross sections directly \cite{spal}. To provide some physical guides, a recent empirical parameterizations for fragments prediction, which is named as the {\sc spacs} \cite{spacs}, for spallation reaction has been adopted to obtain the fragment cross sections. In this work, the BNN approach is employed to reconstruct the residues between the experimental data ($\sigma^{exp}$) and the theoretical predictions ($\sigma^{th}$), i.e.,
\begin{equation}
y_{i}=\mbox{lg}(\sigma^{exp})-\mbox{lg}(\sigma^{th}).
\end{equation}
The cross sections predictions with BNN approach are then given as,
\begin{equation}
\sigma^{BNN+th}=\sigma^{th}\times10^{y^{new}}.
\end{equation}
where $\sigma^{th}$ and $y^{new}$ denotes the {\sc spacs} results and the BNN prediction, respectively. In this work, $\sigma^{th}$ refers to the predictions by the {\sc spacs} parameterizations, which is proposed recently and gains its success in spallation reactions.

\begin{table}[thbp]
\caption{ A list of the adopted data for the measured fragments in the $X$ + p spallation reactions.}
\label{measdata}
\centering
\begin{tabular}{p{48pt}<{\centering}|p{48pt}<{\centering}|p{48pt}<{\centering}|p{48pt}<{\centering}|p{48pt}<{\centering}}
  \hline
  \hline
    $^{A}X$ + $p$  &E(MeV/u) &Numbers &$Z_{i}$ &Reference\\
  \hline
                   &361      &42      &9-17            \\
  $^{36}$Ar + $p$  &546      &42      &9-17      &\cite{LBL-Ar}  \\
                   &765      &38      &9-17               \\
  \hline

  $^{40}$Ar + $p$  &352      &45      &9-17      &\cite{LBL-Ar} \\

  \hline
                    &356   &48      &10-20 \\
  $^{40}$Ca + $p$   &565   &54      &10-20    &\cite{LBL-Ca} \\
                    &763   &54      &10-20    \\
  \hline
                     &300  &128     &10-27      \\
                     &500  &136     &10-27     \\
  $^{56}$Fe + $p$    &750  &148     &8-27    &\cite{56Fe} \\
                     &1000 &152     &8-26 \\
                     &1500 &157     &8-27 \\
  \hline
                     &168  &73      &48-55  &\cite{RIKEN-Xe}   \\
                     &200  &96      &48-55 &\cite{136Xe200} \\
  $^{136}$Xe + $p$   &500  &271     &41-56 &\cite{136Xe500} \\
                     &1000 &604     &3-56 &\cite{136Xe1000} \\
  \hline



  $^{197}$Au + $p$   &800 &352     &60-80  &\cite{197Au800}    \\

  \hline

  $^{208}$Pb + $p$    &500  &249   &69-83  &\cite{208Pb500}    \\
                      &1000 &458   &61-82 &\cite{208Pb1000}    \\

  \hline

  $^{238}$U + $p$     &1000 &364   &74-92   &\cite{238U1000} \\

  \hline
  \hline
\end{tabular}\\
\end{table}

The inputs of neural network are the mass numbers $A_{pi}$ ($A_{i}$), the charge numbers $Z_{pi}$ ($Z_{i}$) of the projectile (fragment) nucleus, and the bombarding energy (in MeV/u) $E_{i}$, i.e., $x_{i} = (A_{pi}, Z_{pi}, E_{i}, Z_{i}, A_{i})$. Systematic experiments have been performed at the Lawernce Berkeley Laboratory (LBL), RI Beam Facility (RIBF) RIKEN, and FRagment Separator (FRS) GSI, which cover a broad range of spallation nuclei from $^{36}$Ar to $^{238}$U. The range of the incident energy changes from 168 MeV/u to 1500 MeV/u, which is relevant for the ADS and proton therapy applications. As listed in Table \ref{measdata}, 3511 data in 20 different reactions will be used in this work. The entire data are divided into two different sets, which serve as the learning set and the validation set, respectively. The learning set is built by randomly selecting 3211 data set and the remaining 300 data as the validation set.

\section{results and discussions}
\label{RSTD}

\begin{figure}
\centering
\includegraphics[width=0.5\textwidth]{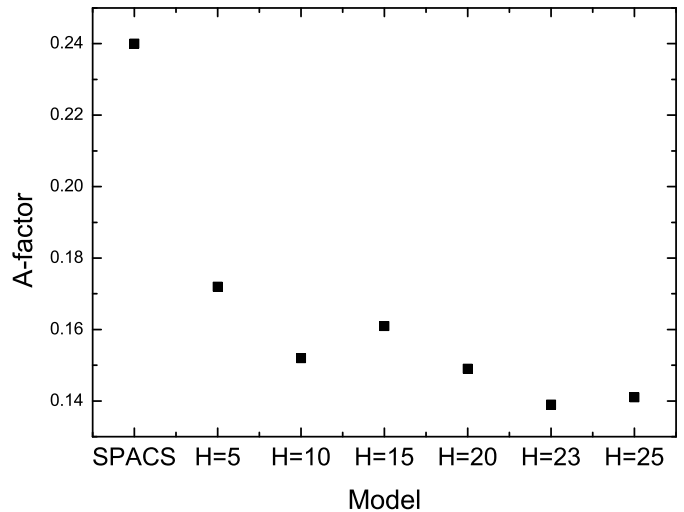}
\caption{The A-factor for the predictions by the {\sc spacs} and the BNN + {\sc spacs} model with different hidden neurons denoted by $H$ of the validation set.}
\label{A-factor}
\end{figure}

The network is trained with different model structures, and 2000 iteration samples are taken in each training. Because the fragment cross sections may differs in several orders of magnitude, an A-factors method \cite{A-factor,AYDS2019} is introduced to indicate the validation results of different models, as shown in Fig. \ref{A-factor}. The A-factor is defined as,
\begin{equation}
A_f=1/N\sum\limits^{N}_{i=1}\frac{|\sigma^{exp} - \sigma^{pre}|}{\sigma^{exp} + \sigma^{pre}},
\end{equation}
where $\sigma^{exp}$ and $\sigma^{pre}$ denotes the measured data and predicted data, respectively. Below we discuss the predictions by the BNN + {\sc spacs} method, and compare them to the measured data and the {\sc spacs} predictions.

In Fig. \ref{A-factor}, the A-factor for the {\sc spacs} predictions and BNN method with different hidden neurons are compared. It is seen that even the BNN with five hidden neurons can significantly improve the predictions. The A-factor decreases with the increasing numbers of $H$. When $H$ is increased to 23, the A-factor cannot be further minimized. A 5-23-1 structure is taken as the optimal network structure, which means that 5 inputs $x_{i} = (A_{pi}, Z_{pi}, E_{i}, Z_{i}, A_{i})$, 1 output $y_{i}=\mbox{lg}(\sigma^{exp})-\mbox{lg}(\sigma^{th})$ and single hidden layer with 23 hidden neurons are included.





\begin{figure*}
\centering
\includegraphics[width=14cm]{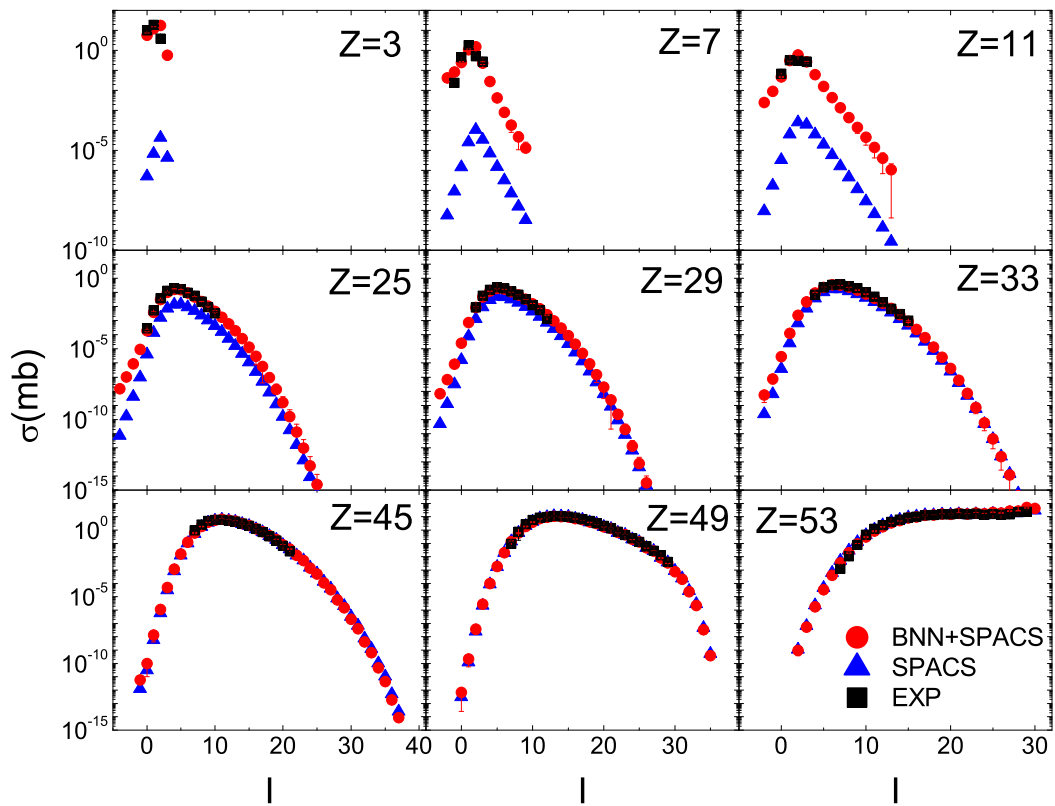}
\caption{The BNN + {\sc spacs} predictions of fragment cross sections of 1000 $A$ MeV $^{136}$Xe + p compared with {\sc spacs} and experimental data (taken from \cite{136Xe1000}). In the $x$ axis $I = N - Z$ denotes the neutron excess of fragment. The measured data, the BNN + {\sc spacs} and the {\sc spacs} predictions are plotted as the squares, circles and triangles, respectively. The experimental and BNN + {\sc spacs} error bars are too small to be shown.}
\label{136Xe1000-I}
\end{figure*}
\begin{figure*}
\centering
\includegraphics[width=14cm]{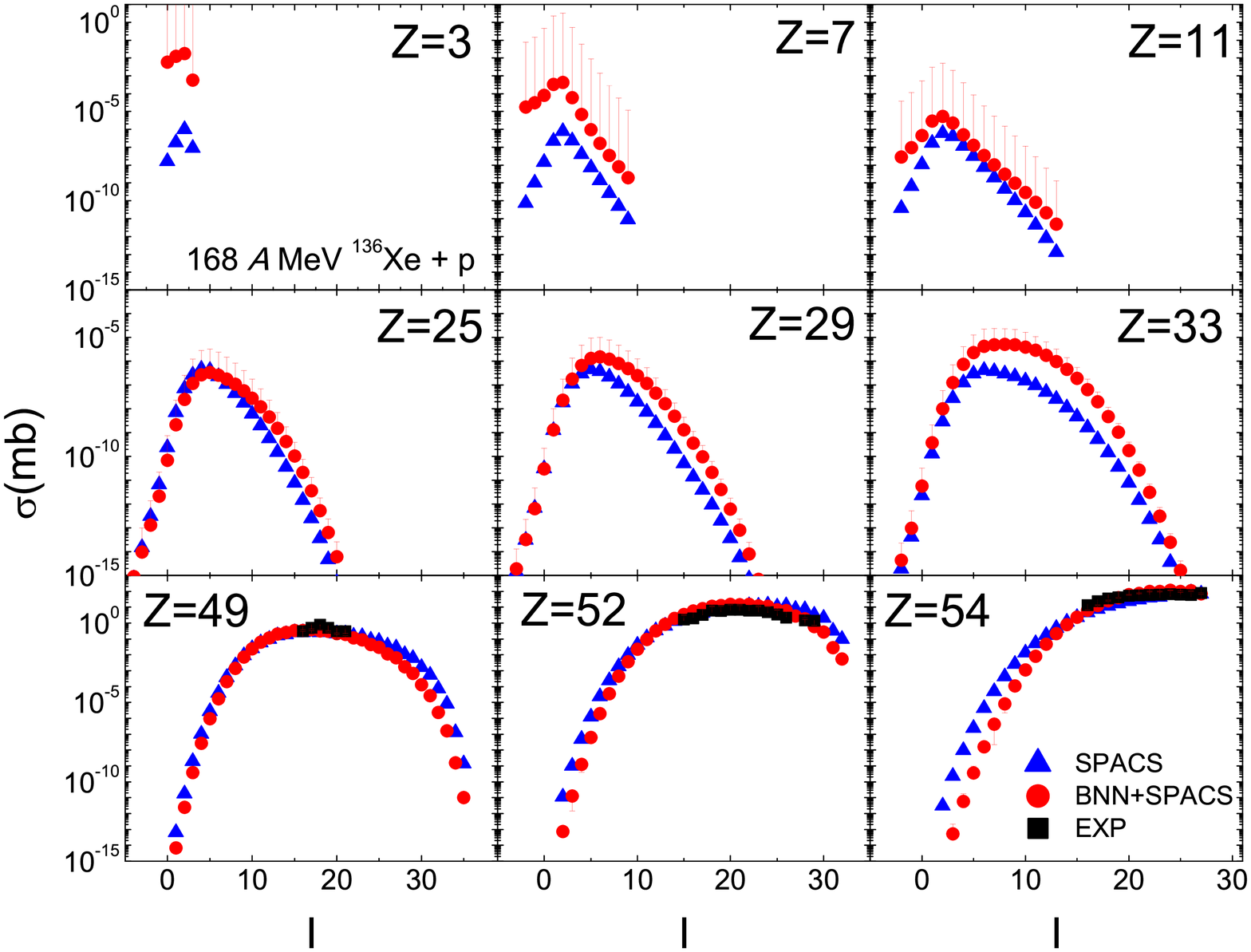}
\caption{Similar to Fig. \ref{136Xe1000-I} but for the 168 $A$ MeV $^{136}$Xe + p reaction (experimental data taken from Ref. \cite{RIKEN-Xe}).}
\label{136Xe168-I}
\end{figure*}
\begin{figure*}
\centering
\includegraphics[width=14cm]{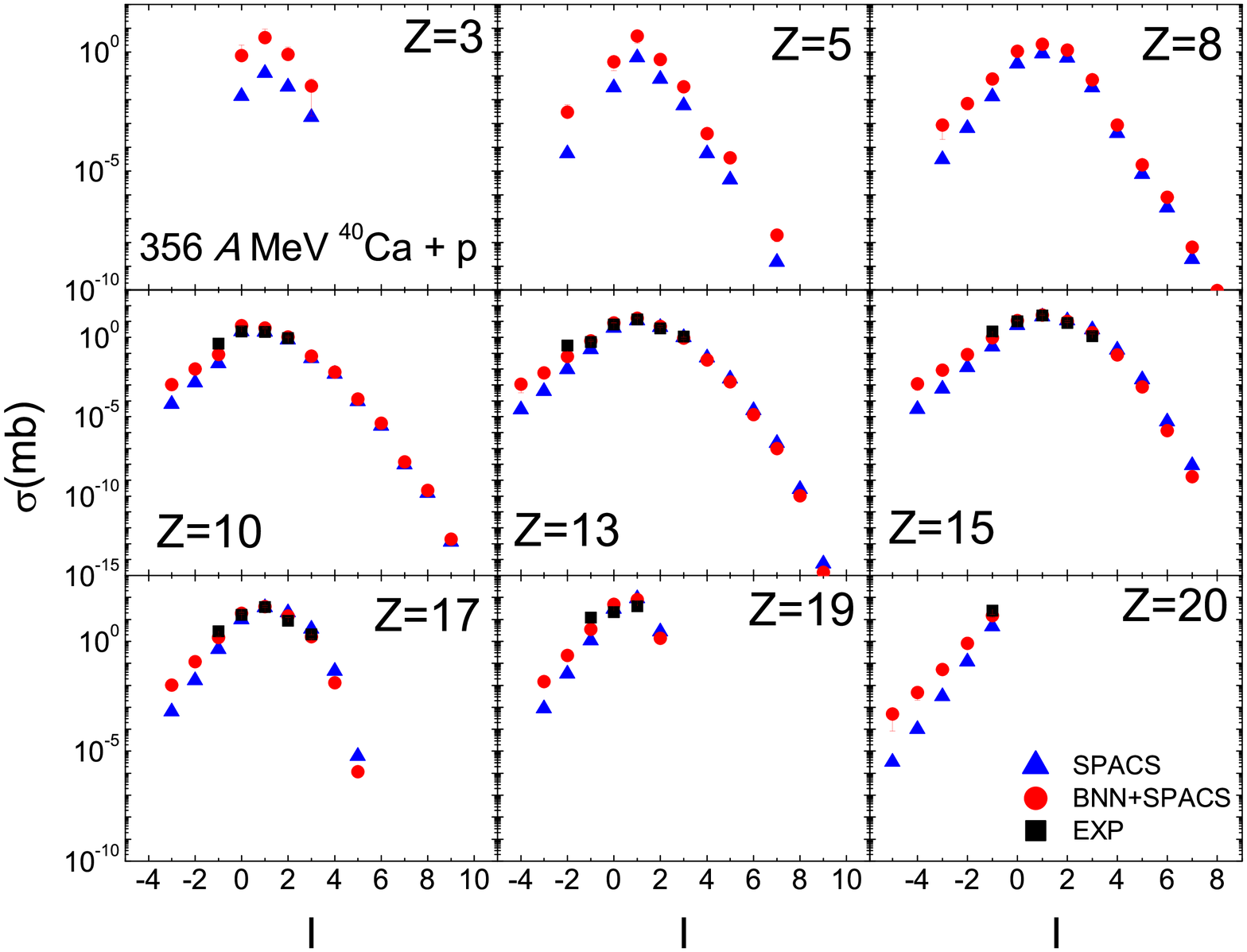}
\caption{Similar to Fig. \ref{136Xe1000-I} but for the 356 $A$ MeV $^{40}$Ca + p reaction (experimental data taken from Ref. \cite{LBL-Ca}).}
\label{40Ca356-I}
\end{figure*}
\begin{figure*}[htbp]
\centering
\subfigure{\includegraphics[width=14cm]{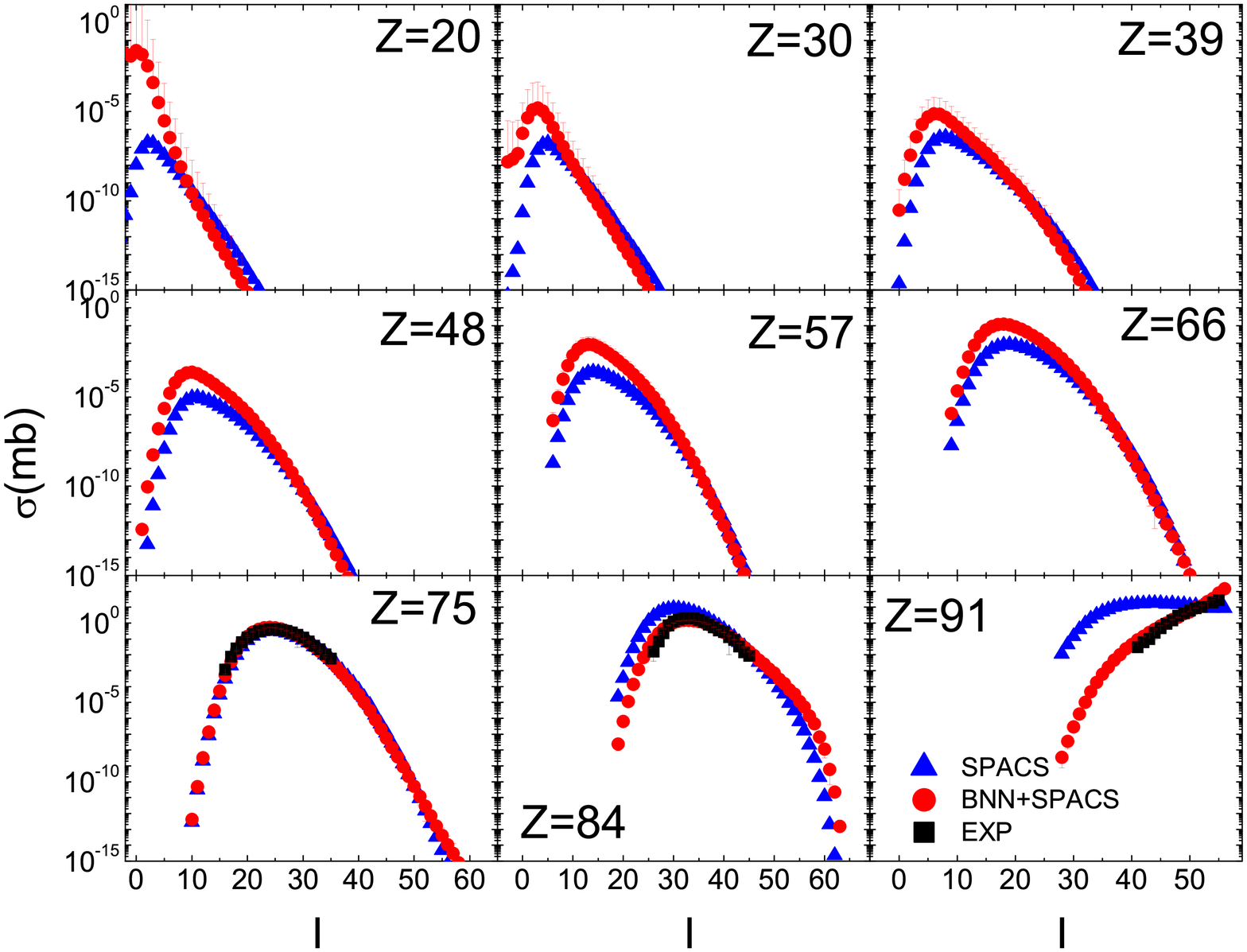}}
\caption{Similar to Fig. \ref{136Xe1000-I} but for the 1 $A$ GeV $^{238}$U + p reaction (experimental data taken from Ref. \cite{238U1000}).}
\label{U2381AGeV}
\end{figure*}

The BNN + {\sc spacs} predictions for fragment cross sections in the 1 $A$ GeV $^{136}$Xe + p, 168 $A$ MeV $^{136}$Xe + p, 356 $A$ MeV $^{40}$Ca + $p$ and 1 $A$ GeV $^{238}$U + $p$ reactions are shown in Fig. \ref{136Xe1000-I} to Fig. \ref{U2381AGeV}, and compared to the experimental data, as well as the {\sc spacs} predictions.

In Fig. \ref{136Xe1000-I}, the predicted and measured  fragment cross sections in the 1 $A$ GeV $^{136}$Xe + p reaction are compared. It is seen that the BNN + {\sc spacs} predictions agree quite well with the measured data for fragments from $Z=$ 3 to $54$. For the fragments with $Z \leq$ 25, the underestimation of experimental data by {\sc spacs} has been improved significantly.

Fig. \ref{136Xe168-I} shows the BNN + {\sc spacs} predictions for fragment cross sections in the 168 $A$ MeV $^{136}$Xe + p reaction, which has been measured at RIBF, RIKEN recently \cite{RIKEN-Xe}. In \cite{RIKEN-Xe}, only the cross sections for the $Z \geq$ 48 fragments are reported. Compared to the measured fragments, the BNN + {\sc spacs} predictions are very close to the {\sc spacs} ones. For the light and the medium fragments ($Z \leq$ 25), the BNN + {\sc spacs} predictions are much higher than the {\sc spacs} ones, which is similar to the results shown in Fig. \ref{136Xe1000-I}. In addition, the uncertainties are relatively large for the $Z \leq$ 11 isotopes, which may be caused by the insufficient data in the training set in this incident energy.

The spallation of intermediate nuclei are of interests in the proton therapy and nuclear astrophysics. The composition of interstellar  matter, which are influenced by the cosmic ray (mainly high energy proton) induced spallation reactions. The $^{40}$Ca + p at 356 $A$ MeV have been studied, for which the predicted and measured results are shown in Fig. \ref{40Ca356-I}. Compared to the measured results, the BNN + {\sc spacs} and {\sc spacs} predictions both can reproduce the experimental data quiet well. The predicted fragment cross sections by the BNN + {\sc spacs} method are in line with those by {\sc spacs} except for the $Z =$ 3 isotopes.

The predictions to the fragment cross sections in the 1 $A$ GeV $^{238}$U + p reaction are compared in Fig. \ref{U2381AGeV}. The measured data cover the fragments from $Z =$ 74 to 92 \cite{238U1000}. It can be seen that the BNN + {\sc spacs} method can predict the results well, while the {\sc spacs} highly overestimate the measured results for $Z=$ 91 and 92. The BNN + {\sc spacs} method show the sign of larger than the {\sc spacs} method for the fragments of smaller $I$. It seems that for the $Z =$ 20 isotopes, fragments For the spallation of a heavy system as $^{238}$U, the predictions by BNN + {\sc spacs} become worse, which indicates that the BNN should be further improved by incorporating more data for small $Z$ fragments produced in the heavy systems.

{\begin{figure*}[htbp]
\centering
\subfigure{\includegraphics[width=8cm]{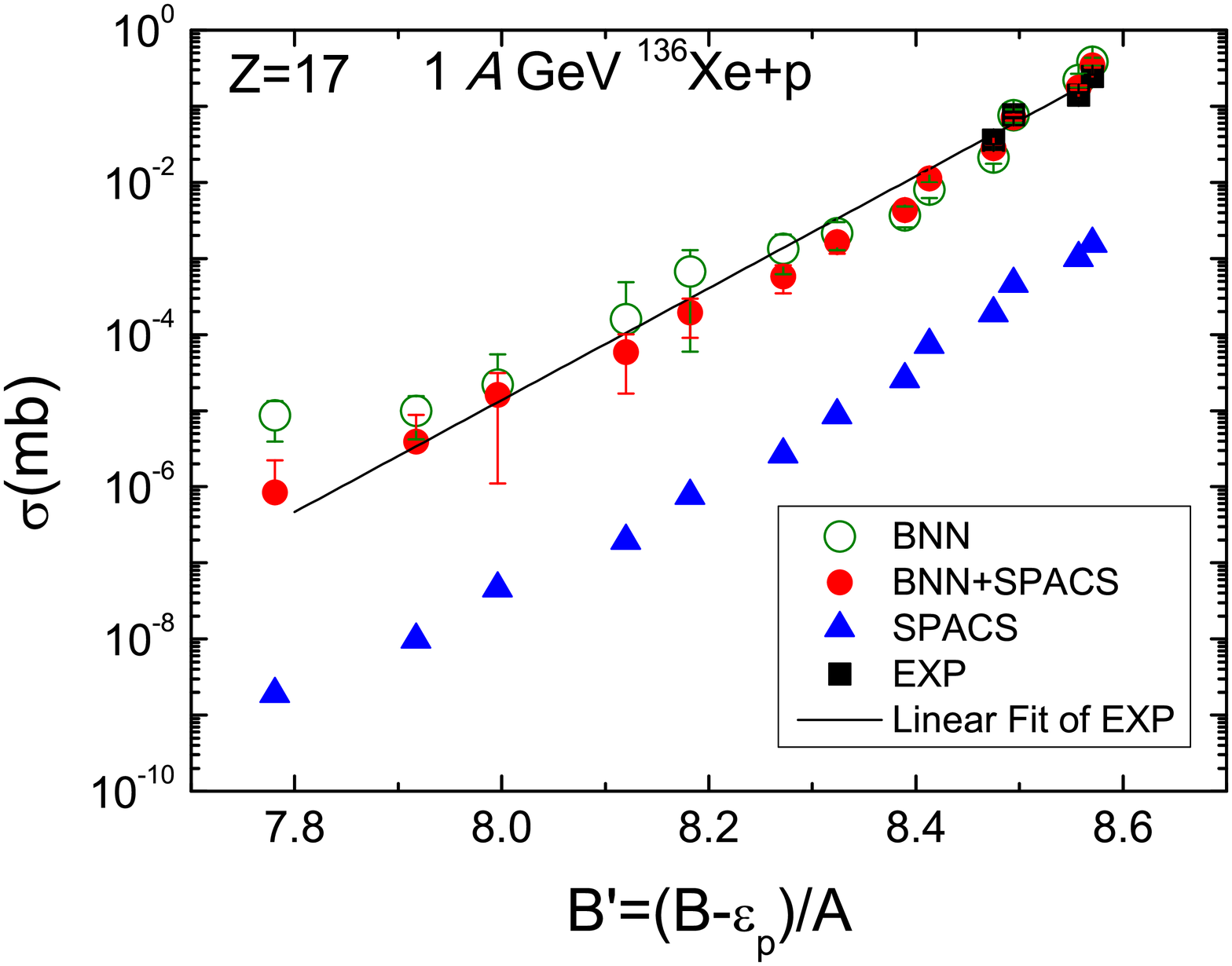}}
\subfigure{\includegraphics[width=8cm]{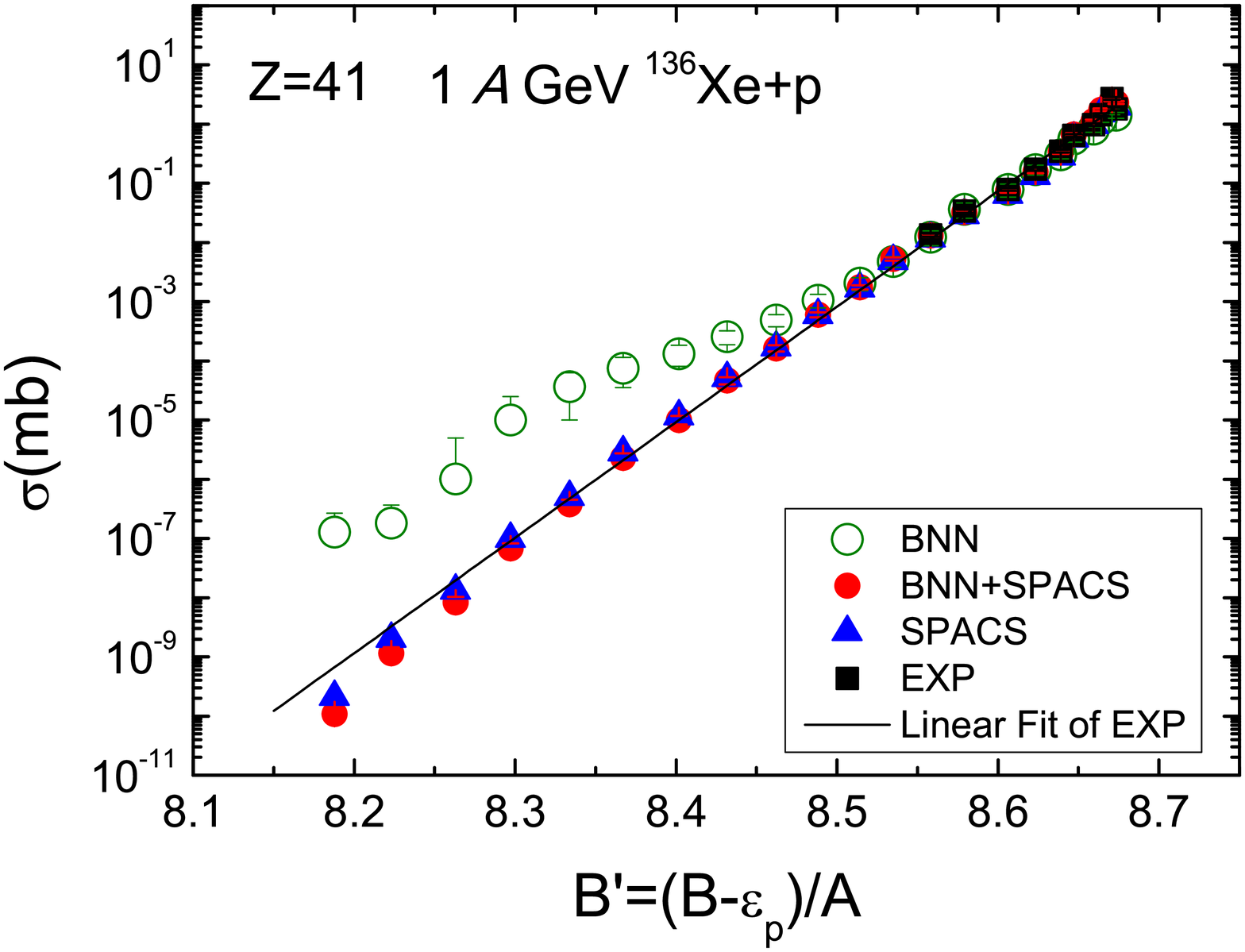}}
\caption{Isotopic cross section dependence on average binding energy for fragments of $Z =$ 17 and 41 produced in the 1 $A$ GeV $^{136}$Xe + p reaction (experimental data taken from \cite{136Xe1000}). The open circles, solid circles, triangles and squares denote the data by the BNN predictions (see Ref. \cite{spal}), BNN + {\sc spacs} in this work, {\sc spacs}, and the measured data, respectively. The solid line denote the fitting to the experimental data (see text for explanation).}
\label{CRS-B}
\end{figure*}

The BNN + {\sc spacs} predictions is further verified by using the correlation between the cross section and average binding energy, which has been performed in Ref. \cite{spal}. It is generally believed that the isotopic cross section depends on the average binding energy in the form of $\sigma= C \mbox{e}^{(B'-8)/\tau}$, where $C$ and $\tau$ are free parameters, and ${B'=(B-\epsilon_{p})/A}$ (in which $\epsilon_{p}=0.5[(-1)^{N}+(-1)^{Z}]\epsilon_{0} A^{-3/4}$ is the pairing energy for the fragment, and $\epsilon_{0}=$ 30 MeV). It is clearly seen that the predicted isotopic cross sections by BNN + {\sc spacs} model for the $Z=$ 17 and 41 obey the correlation very well, which improves both the previous BNN method and also the {\sc spacs} method.

From the above results, which cover the fragment cross section predictions for proton induced reactions from the intermediate nuclei to the heavy nuclei, and for the incident energy from 168 $A$ MeV to 1 $A$ GeV, it is seen that the BNN approach improves the quality of the empirical {\sc spacs} parameterizations through the reconstruction of the residual cross sections between the {\sc spacs} predictions and measured data. But the BNN + {\sc spacs} method can be a new tool to predict the fragment cross section in the spallation reactions since it can work independently after the network is formed.

If we revisit the foundation of the BNN approach, it is natural that the BNN + {\sc spacs} model should have a better prediction than the {\sc spacs} parameterizations since the difference between the {\sc spacs} and the measured data has been minimized. This is why the BNN + {\sc spacs} improves the prediction, and also avoids the nonphysical phenomenon by forming a direct BNN learning network from the measured data as shown in Ref. \cite{spal}. The physical implantations of {\sc spacs} play important roles to make the BNN + {\sc spacs} method reasonable in physics, and the leaning and predicting abilities of the BNN also improve the predictions where the {\sc spacs} parameterizations do not work well.

It is indicated from the results that the {\sc spacs} tends to underestimate the cross sections for fragments with relative small $Z$, while overestimates the fragments with $Z$ close to the heavy spallation nuclei. These shortcomings have been overcome by the BNN + {\sc spacs} model. Limitations still exist for the constructed BNN + {\sc spacs} model in this work since the absence of experimental data for reactions of incident energy below 100 $A$ MeV, and the small spallation systems. For the applications in the proton therapy, the incident energy maybe lower than 100 $A$ MeV, and the mass of the spallation nuclei are smaller than 30. The smallest spallation reaction adopted in this work is for $^{36}$Ar. If we consider the interstellar matters, most of the nuclei have mass numbers smaller than 56. In spallation reactions induced by the high energy cosmic rays and in the proton therapy process, we should improve the prediction model to cover the small spallation systems, for which the {\sc spacs} parameterizations do not work well. It is important to improve the BNN + {\sc spacs} predictions by introducing new data for the spallation reactions of intermediate energy (for example below 100 $A$ MeV) and intermediate/small systems ($A < $ 20), which calls for new experiments.

\section{Conclusion}
\label{CONCLD}
In this article, the BNN approach is proposed to predict the fragment cross sections in proton induced spallation reactions combined to the {\sc spacs} parameterizations. Based on the 3,511 measured fragment cross sections in 20 spallation reaction systems, the optimal network structure has been established to be 5-23-1, which includes 5 inputs, 1 output and a single hidden layer with 23 hidden neurons. By reconstructing the residuals between the measured data and the {\sc spacs} predictions, the  BNN + {\sc spacs} method is verified to well reproduce the experimental data. It is also shown that the BNN + {\sc spacs} method can yield a better global prediction compared to the {\sc spacs} parameterizations. The established BNN + {\sc spacs} method potentially can be applied into the researches of nuclear physics, nuclear astrophysics, ADS, and proton therapy, etc.

\section*{Acknowledgement}
This work is supported by the National Natural Science Foundation of China (grant Nos. U1732135 and 11975091).

\end{document}